## 12.2 Muon Collider

S.D. Holmes, V.D. Shiltsev

Both $e^+e^-$ and $\mu^+\mu^-$ colliders have been proposed as possible candidates for a lepton collider to complement and extend the reach of the Large Hadron Collider (LHC) at CERN. The physics program that could be pursued by a new lepton collider ($e^+e^-$ or $\mu^+\mu^-$) with sufficient luminosity would include understanding the mechanism behind mass generation and electroweak symmetry breaking; searching for, and possibly discovering, supersymmetric particles; and hunting for signs of extra spacetime dimensions and quantum gravity. However, the appropriate energy reach for such a collider is currently unknown, and will only be determined following initial physics results at the LHC. It is entirely possible that such results will indicate that a lepton collider with a collision energy well in excess of 1 TeV will be required to illuminate the physics uncovered at LHC. Such a requirement would require consideration of muons as the lepton of choice for such a collider.

The lifetime of the muon, 2 μs in the muon rest frame, is just long enough to allow acceleration to high energy before the muon decays into an electron, a muon-type neutrino and an electron-type antineutrino ($\mu^- \to e^- \nu_\mu \bar{\nu}_e$). However, constructing and operating a muon based collider with useable luminosity requires surmounting significant technical challenges associated with the production, capture, cooling, acceleration, and storage of muons in the required quantities and with appropriate phase space densitites. Over the last decade there has been significant progress in developing the concepts and technologies needed to produce, capture, cool, and accelerate muon beams with high intensities of the order of $O(10^{21})$ muons/year. These developments have established a multi-TeV Muon Collider (MC) in which $\mu^+$ and $\mu^-$ are brought to collision at high luminosity in a storage ring as a viable option for the next generation lepton-lepton collider for the full exploration of high energy physics in the era following the LHC discoveries.

Muon colliders were proposed by Budker [1] in 1969 and later conceptually developed by a number of authors and collaborations (see comprehensive list of references in [2]). Fig. 12.3 presents a possible layout on the Fermilab site of a MC that would fully explore the physics responsible for electroweak symmetry breaking. Such a MC requires a center-of-mass energy ($\sqrt{s}$) of a few TeV and a luminosity in the $10^{34}$ cm$^{-2}$s$^{-1}$ range (see Table 12.2 for the list of parameters). The MC consists of a high power proton driver based, e.g., on the "Project X" SRF-based 8 GeV 2-4 MW H$^-$ linac [3]; pre-target accumulation and compressor rings where very high intensity 1-3 ns long proton bunches are formed; a liquid mercury target for converting the proton beam into a tertiary muon beam with energy of about 200 MeV; a multi-stage ionization cooling section that reduces the transverse and longitudinal emittances and creates a low emittance beam; a multistage acceleration (initial and main) system – the latter employing Recirculating Linear Accelerators (RLA) to accelerate muons in a modest number of turns up to 2 TeV using superconducting RF technology; and, finally, a roughly 2-km diameter Collider Ring located some 100 meters underground where counter-propagating muon beams are stored and collide over the roughly 1000-2000 turns corresponding to the muon lifetime.

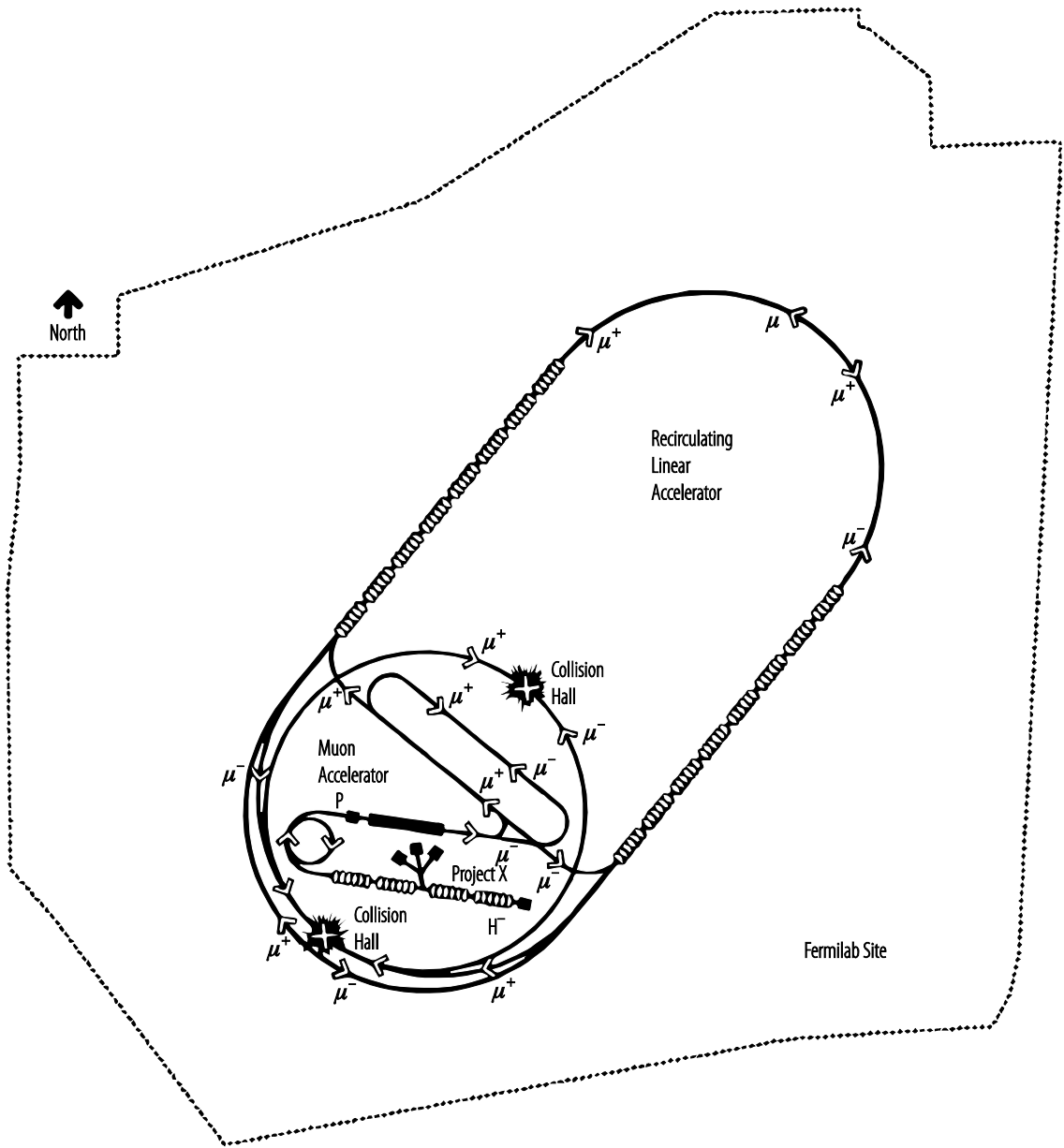

**Fig. 12.3.** Schematic of a 4 TeV Muon Collider on the 6×7 km FNAL site.

## 12.2.1 Technical Motivations

Synchrotron radiation (proportional to the fourth power of the Lorentz factor $\gamma^4$) poses severe limitations on multi-TeV $e^+e^-$ colliders, namely they must have a linear, not circular, geometry. Practical acceleration schemes then require a facility tens of kilometers long. Furthermore, beam-beam effects at the collision point induce the electrons and positrons to radiate, which broadens the colliding beam energy distributions. Since $(m_\mu/m_e)^4 = 2\times10^9$, all of these radiation-related effects can be mitigated by using muons instead of electrons. A multi-TeV $\mu^+\mu^-$ collider can be circular and therefore have a compact geometry that will fit on existing accelerator sites, and may be significantly less expensive than alternative machines. The center-of-mass energy spread for a 3-TeV $\mu^+\mu^-$ collider, $dE/E < 0.1\%$, is an order of magnitude smaller than for an $e^+e^-$ collider of the same energy. Additionally, the MC needs lower wall plug power and has a smaller number of elements requiring high reliability and individual control for effective operation [4].

An additional attraction of a MC is its possible synergy with the Neutrino Factory concept [5]. The front-end of a MC, up to and including the initial cooling channel, is similar (perhaps identical) to the corresponding Neutrino Factory (NF) front-end [6]. However, in a NF the cooling channel must reduce the transverse emittances ($\varepsilon_x$, $\varepsilon_y$) by only factors of a few, whereas to produce the desired luminosity, a MC cooling channel must reduce the transverse emittances (vertical and horizontal) by factors of a few hundred and reduce the longitudinal emittance $\varepsilon_L$ by a factor $O(10)$. Thus, a Neutrino Factory could offer the opportunity of a staged approach to a Muon Collider, and also the opportunity of shared R&D.

## 12.2.2 Design Concepts

Since muons decay quickly, large numbers of them must be produced to operate a muon collider at high luminosity. Collection of muons from the decay of pions produced in proton-nucleus interactions results in a large initial phase volume for the muons, which must be reduced (cooled) by a factor of $10^6$ for a practical collider. Without such a cooling, the luminosity reach will not exceed $O(10^{31}\ \mathrm{cm^{-2}s^{-1}})$, a substantial limitation on the discovery reach of the MC. The technique of ionization cooling [7] is proposed for the $\mu^+\mu^-$ collider [8,9]. This technique is uniquely applicable to muons because of their minimal interaction with matter.

Ionization cooling involves passing the beam through some material absorber in which the muons lose momentum essentially along the direction of motion via ionization energy loss, commonly referred to as $dE/dx$. Both transverse and longitudinal momentum are reduced via this mechanism, but only the longitudinal momentum is then restored by reacceleration, leaving a net loss of transverse momentum (transverse cooling). The process is repeated many times to achieve a large cooling factor. The energy spread can be reduced by introducing a transverse variation in the absorber density or thickness (e.g., a wedge) at a location where there is dispersion (a correlation between transverse position and energy). This method results in a corresponding increase of transverse phase space and represents in an exchange of longitudinal and transverse emittances. With transverse cooling, this allows cooling in all dimensions. The cooling effect on the emittance is balanced against stochastic multiple scattering and Landau straggling, leading to an equilibrium emittance.

Theoretical studies have shown that, assuming realistic parameters for the cooling hardware, ionization cooling can be expected to reduce the phase space volume occupied by the initial muon beam by a factor of $10^5$–$10^6$. A complete cooling channel would consist of 20–30 cooling stages, each stage yielding about a factor of 2 in 6D phase space reduction – see Fig. 12.4.

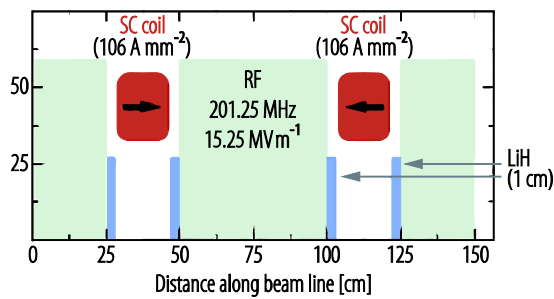

**Fig. 12.4.** Cooling-channel section. Muons lose energy in lithium hydride (LiH) absorbers (*blue*) that is replaced when the muons are reaccelerated in the longitudinal direction in radio frequency (RF) cavities (*green*). The few-Tesla superconducting (SC) solenoids (*red*) confine the beam within the channel and radially focus the beam at the absorbers. Some representative component parameters are also shown (from Ref. [2]).

Such a cooling method seems relatively straightforward in principle, but has proven quite challenging to implement in practice. One of the main issues is breakdown suppression and attainment of high accelerating gradients in normal-conducting RF cavities immersed in strong magnetic fields. The International Muon Ionization Cooling Experiment (MICE [10]) at RAL (UK) was set to test an ionization cooling channel cell consisting of a sequence of LiH absorbers and 201 MHz rf cavities within a lattice of solenoids that provide the required focusing in a 200 MeV muon beam [11]. The initial results indicate anticipated significant emittance reduction $O(10\%)$ in the "no re-acceleration" configuration [12] and, therefore, can be considered as the first experimental proof of the ionization cooling concept.

### 12.2.3 Technology Development

Multi-MW target R&D has greatly advanced in recent years, and has culminated in the Mercury Intense Target experiment (MERIT [13]) which has successfully demonstrated a Hg-jet injected into a 15T solenoid and hit by an intense proton beam from the CERN PS. A high-$Z$ target is chosen to maximize $\pi^{\pm}$ production. The solenoid radially confines essentially all the $\pi^{\pm}$ coming from the target. The Hg-jet choice avoids the shock and radiation damage related target-lifetime issues that arise in a solid target. The jet was viewed by high speed cameras which enabled measurement of the jet dynamics. MERIT results suggest this technology could support beam powers in excess of 4 MW. More advanced solutions for multi-MW targets are under considerations, too, such as granular waterfall targets [14].

Significant efforts are presently focused on high gradient normal conducting rf cavities operating in multi-Tesla magnetic fields as required in the bunching, phase rotation, and cooling channel designs. Closed 805MHz rf cells with thin Be windows have initially shown significant reduction of maximum rf gradient in a 3T field – 12MV/m vs. 17MV/m specified. Further R&D as part of the U.S. based Muon Accelerator Program (MAP) has experimentally demonstrated some 50 MV/m gradients in the rf cavities with high pressure hydrogen gas [15] and in the Be-coated vacuum cavities [16].

Several self-consistent concepts based on different technologies have recently emerged for the MC 6-dimensional cooling channel which plays a central role in reaching high luminosity. To achieve the desired mixing of transverse and longitudinal degrees of freedom, the muons must either pass through a series of wedge absorbers in a ring [17] or be put onto a helical trajectory, e.g., as in a "Helical Cooling Channel" [18] or a "FOFO-snake" [19]. The design simulations of the channels are not yet complete and the main challenges are attainment of sufficiently large dynamic apertures, taking into account realistic magnetic fields, RF cavities and absorbers, optimization of the B-fields in RF cavities and technological complexity. The design of the final cooling stages is particularly challenging as it requires very high solenoid fields (up to ~30T have been considered [20]). The final MC luminosity is proportional to this field. High-field superconducting magnets for the collider ring and for the cooling have been actively developed [21], including feasibility studies of a high temperature superconductor (HTS) option for the 25-50 T final cooling solenoids [22].

A Recirculating Linac with SC RF cavities (e.g. 1.3 GHz ILC-like cavities) is a very attractive option for acceleration of muons from the low energies emerging from the cooling sections to the energy of the experiments. The recirculating linac offers small lengths and low wall plug power consumption but requires small beam emittances.

Recently, realistic collider ring beam optics has been designed which boasts a very good dynamic aperture for about $dP/P = \pm 0.5\%$ and small momentum compaction [23, 24]. The distortions due to the beam-beam interaction will need to be studied as well as practical issues of the machine-detector interface.

Representative peformance parameters for a multi-TeV Muon Collider are given in Table 12.2. These parameters are based on the design concepts described above and represent reasonable extrapolations of technologies currently under development. The luminosities displayed are appropriate for the physics research programs that would be undertaken at such a facility [25-28].

**Table 12.2.** The parameters of the low- and high-energy Muon Collider options.

| Parameter | Higgs Factory | Low $E$ | High $E$ |
|---|---|---|---|
| Center-of-mass energy [TeV] | 0.126 | 1.5 | 6 |
| Luminosity [cm$^{-2}$s$^{-1}$] | 0.005·10$^{34}$ | 4.5·10$^{34}$ | 7·10$^{34}$ |
| Number of bunches | 1 | 1 | 1 |
| Muons/bunch [10$^{12}$] | 2 | 2 | 2 |
| Circumference [km] | 0.3 | 2.8 | 6.3 |
| Focusing at IP $\beta^*$ / $\sigma_z$ [mm] | 25/5 | 10/10 | 10/5 |
| Beam energy spread $dp/p$ (rms) [%] | 0.003 | 0.1 | 0.10 |
| Ring depth [m] | ~10 | 13 | ~150 |
| Proton driver pulse rate [Hz] | 30 | 12 | 15 |
| Proton driver power [MW] | ≈4 | ≈4 | ≈2 |
| Transverse emittance $\varepsilon_T$ [π μmrad] | 300 | 25 | 25 |
| Longitudinal emittance $\varepsilon_L$ [π mmrad] | 1 | 72 | 72 |

## 12.2.4 Advanced Muon Collider Concepts

In the last few years several advanced muon collider concepts were proposed. An alternative low emittance muon source based on near-threshold production of muons in the reaction e+e- → μ+ μ-was considered in Ref. [29]. The scheme relies on availability of high intensity beam of 45 GeV positrons hitting solid, liquid or crystal target of Be, C or diamond. The resulting emittance of the muon beam is very small and allows direct acceleration with extensive ionization cooling. Synchrotron radiation of high-energy muons channelling in between crystal planes results in very small emittances, too, and opens opportunities for crystal-based muon colliders. Given natural advantages of muons, such as absence of nuclear interaction characteristic of protons and greatly reduced synchrotron radiation compared to electrons, the muons are particle of choice for ultra-high gradient acceleration in crystals, originally proposed in [30]. Such colliders with gradients $O(0.1-1$ TeV/m$)$ can potentially reach c.o.m energies hundreds of times higher than in the LHC collisions, though, by necessity, with lower luminosities due to practical limits on the facility total electrical power consumption $O(100MW)$ [31]. Of course, significant R&D is needed to demonstrate feasibility of the channelling acceleration in crystals or, as a first step, in carbon nanotubes [32].

## References for 12.2